\documentclass[doublecol]{epl2}

\title{Quantum Dot in a Hybrid Structure with Dipolar Excitons}
\shorttitle{Quantum Dot in a Hybrid Structure with Dipolar Excitons} 

\author{M.M. Mahmoodian\inst{1,2} \and A.V. Chaplik\inst{1,2} }
\shortauthor{M.M. Mahmoodian and A.V. Chaplik}

\institute{
  \inst{1} Rzhanov Institute of Semiconductor Physics, Siberian Branch of the Russian Academy of Sciences - Novosibirsk, 630090, Russia\\
  \inst{2} Novosibirsk State University - Novosibirsk, 630090, Russia
}

\abstract{Electron states in a quantum dot (QD) located near a 2D system of dipolar excitons are perturbed by fluctuations of the exciton density caused by the electron-exciton interaction. This results in the frequency changes of electron transitions in a QD. The frequency depends on the exciton density, as well as on the exciton gas phase state. In the present work, the shifts of the two lowest QD energy levels are found both in the normal state of the exciton system and for the Bose-Einstein condensation (BEC) regime.}

\begin{document}

\maketitle

\section{Introduction}
Two-dimensional (2D) systems of excitons and exciton-polaritons have been studied for a rather long time. The most impressive effect in such systems is the one resembling the BEC phase transition: the exciton recombination line shape drastically changes and this indicates the emergence of a new phase in the system \cite{a1,a2,a3,a4,a5,a6}. More varied possibilities for researches are available in hybrid electron-exciton structures, especially in the case of spatially indirect dipolar excitons with a long life-time. In the papers \cite{a7,a8,a9,a10,a11,a12,a13,a14,a15} such hybrid structures were considered for spatially uniform 2D electron gas. We guess that electron nanostructures as one of components of such hybrid systems are also of considerable interest. If, for example, a set of QDs is placed close to the 2D gas of dipolar excitons, then the interaction of electrons in QDs with excitons results in the shift (splitting) of the QDs energy levels observable in the optical spectra. Evidently, the results will be different for the normal Bose-gas of excitons and for the BEC phase. Thus, there appears an additional possibility to investigate the phase transition in the exciton system by means of its influence on the properties of electron component in a hybrid structure. In particular it would be rather interesting to obtain data on the density and effective coupling constant of the dipole gas from the optical experiments with QDs. For example, the Raman spectroscopy of individual QDs (free of the inhomogeneous broadening) demonstrates a rather high resolution at the level of $10^{-5}$ eV (see below).

In the present paper, we consider the two abovementioned regimes of QD interaction with the 2D gas of dipolar excitons.

\section{Imperfect Bose-gas of excitons at $T=0$}

\begin{figure}
\onefigure[width=5.5cm]{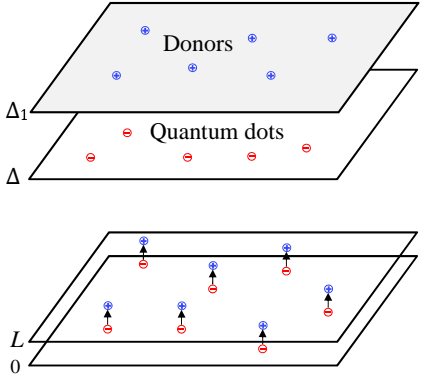}
\caption{Schematic representation of the hybrid structure.}
\label{fig1}
\end{figure}

We describe the exciton condensate by the Gross-Pitaevskii equation. Its applicability condition is detailed in the review by Pitaevskii \cite{a16}. In our case, the external potential in this equation should be replaced by the QD interaction with excitons. Denoting the QD electron wave function by $\chi$ and the condensate one - by $\psi$ we have the electron-exciton interaction energy in the form:
\begin{eqnarray}\label{f1}
\hat{H}_{int}=\int|\psi(\bm{\rho})|^2V_{\mbox{e-ex}}(\bm{\rho}-\bm{\rho}')|\chi(\bm{\rho'})|^2d\bm{\rho}d\bm{\rho}',
\end{eqnarray}
where $V_{\mbox{e-ex}}$ is the pair electron-exciton potential. The function $V_{\mbox{e-ex}}(\bm{\rho}-\bm{\rho}')$ can be found from a simple electrostatic problem (see Fig.~\ref{fig1}): plane $z=0$ is occupied by dipoles with charges $\pm e$ and the shoulder $L$ oriented along the $z$-axis. The QD is placed at the distance $\Delta\gg L$ from this plane and the QD height is much less than its sizes in the $x$-$y$-plane; thus, electron states in a QD correspond to the ultraquantum limit of the transversal motion. Then we obtain:
\begin{eqnarray}\label{f2}
V_{\mbox{e-ex}}(\bm{\rho}-\bm{\rho}')=-\frac{\tilde{e}^2\Delta L}{\left[\left(\bm{\rho}-\bm{\rho}'\right)^2+\Delta^2\right]^{3/2}},
\end{eqnarray}
$\bm{\rho},\bm{\rho}'$ are the 2D radius-vectors in the $x$-$y$-plane. From here and further in what follows, $\tilde{e}^2$ denotes $e^2/\varepsilon$, where $\varepsilon$ is the background permeability; the sign in the right side of eq.~(\ref{f2}) corresponds to a certain polarity of dipoles, i.e. attraction to an electron.

Then one has to account for the Coulomb field of the charge that compensates the QD electron charge. To be certain, we consider the structure shown in Fig.~\ref{fig1}, where the QDs are populated with electrons due to the $\delta$-doping. Donors are separated from the QD plane by the tunnel-transparent barrier. Such structures with $Ge$ QDs on the $Si$ surface were reported in \cite{a17}. A single-ionized donor interaction with an exciton is given by the formula: $V_{\mbox{D-ex}}=\tilde{e}^2L\Delta_1/(\bm{\rho}^2+\Delta_1^2)^{3/2}$ (exciton coordinates are $\bm{\rho}$, $z=0$, and donor coordinates - $\bm{\rho}=0$, $z=\Delta_1$).

As for the exciton-exciton interaction $V_{\mbox{ex-ex}}$, it can be treated as a contact one: $V_{\mbox{ex-ex}}=g\delta(\bm{\rho}-\bm{\rho}')$ with $g=4\pi\tilde{e}^2L$ because the repulsion of parallel dipoles decreases at a large separation as $|\bm{\rho}-\bm{\rho}'|^{-3}$ that is a short-range potential in the 2D case.

Finally, one has also to account for $U_0(\bm{\rho})$ - the potentials of QD and of the remote donor which keep an electron within a QD. Summing up all contributions to the energy of the system we come to the self-consistent equations for functions $\psi$ and $\chi$ ($\hbar=1$):
\begin{eqnarray}\label{f3}
&&\bigg[-\frac{1}{2M}\Delta_{\bm{\rho}}-\mu+g|\psi(\bm{\rho})|^2+\\
&&+\int V_{\mbox{e-ex}}(\bm{\rho}-\bm{\rho}_1)|\chi(\bm{\rho}_1)|^2d\bm{\rho}_1+V_{\mbox{D-ex}}(\bm{\rho})\bigg]\psi(\bm{\rho})=0,\nonumber\\
&&\bigg[-\frac{1}{2m}\Delta_{\bm{\rho}'}+\int V_{\mbox{e-ex}}(\bm{\rho}'-\bm{\rho}_1)|\psi(\bm{\rho}_1)|^2d\bm{\rho}_1+\nonumber\\
&&+U_0(\bm{\rho}')-E\bigg]\chi(\bm{\rho}')=0.\nonumber
\end{eqnarray}

Here, $M$ and $m$ are the exciton and electron masses, respectively, $\mu$ - the exciton gas chemical potential, $E$ - electron energy in a QD.

In the absence of interactions $V_{\mbox{e-ex}}$ and $V_{\mbox{D-ex}}$ the solution  for $\psi(\bm{\rho})$ in eq.~(\ref{f3}) is simply constant: $\psi=\sqrt{n_0}$, where $n_0$ is the uniform (unperturbed) exciton gas density and the equation itself gives the chemical potential value $\mu=gn_0$.

Interaction with electron and ionized donor results in the exciton condensate density fluctuation: $\psi=\sqrt{n_0}+\varphi(\bm{\rho})$. We suppose the fluctuation is small, as compared with the initial density $n_0$ and linearize system (\ref{f3}) by $\varphi$. The criterion for the applicability of this approximation will be expressed below in terms of the characteristic parameters of the structure.

After the linearization we get system (\ref{f3}) in the form:
\begin{eqnarray}
&&-\frac{1}{2M}\Delta_{\bm{\rho}}\varphi(\bm{\rho})+2gn_0\varphi(\bm{\rho})+\frac{\tilde{e}^2L\Delta_1\sqrt{n_0}}{\left(\bm{\rho}^2+\Delta_1^2\right)^{3/2}}-\nonumber\\
&&-\tilde{e}^2\Delta L\sqrt{n_0}\int\frac{\left|\chi(\bm{\rho}_1)\right|^2d\bm{\rho}_1}{\left[(\bm{\rho}-\bm{\rho}_1)^2+\Delta^2\right]^{3/2}}=0,\label{f4}\\
&&-\frac{1}{2m}\Delta_{\bm{\rho}'}\chi(\bm{\rho}')-\left(E+2\pi\tilde{e}^2Ln_0\right)\chi(\bm{\rho}')+U_0(\bm{\rho}')\chi(\bm{\rho}')-\nonumber\\
&&-2\tilde{e}^2\Delta L\sqrt{n_0}\int\frac{\varphi(\bm{\rho}_2)d\bm{\rho}_2}{\left[(\bm{\rho}'-\bm{\rho}_2)^2+\Delta^2\right]^{3/2}}\chi(\bm{\rho}')=0.\label{f5}
\end{eqnarray}

The formal solution of eq.~(\ref{f4}) can be given by making use of its Green function $G(\bm{\rho},\bm{\rho}')=(1/2\pi)K_0(\kappa|\bm{\rho}-\bm{\rho}'|)$, where $\kappa^2=4Mgn_0$, $K_0$ is the McDonald function:
\begin{eqnarray}\label{f6}
&&\varphi(\bm{\rho})=-\frac{M\tilde{e}^2\Delta_1 Ln_0}{\pi}\int\frac{K_0(\kappa|\bm{\rho}-\bm{\rho}_1|)d\bm{\rho}_1}{\left[\bm{\rho}_1^2+\Delta_1^2\right]^{3/2}}+\\
&&+\frac{M\tilde{e}^2\Delta L\sqrt{n_0}}{\pi}\int\frac{K_0(\kappa|\bm{\rho}-\bm{\rho}_1|)\left|\chi(\bm{\rho}_2)\right|^2d\bm{\rho}_1d\bm{\rho}_2}{\left[(\bm{\rho}_1-\bm{\rho}_2)^2+\Delta^2\right]^{3/2}}.\nonumber
\end{eqnarray}

The substitution of $\varphi(\bm{\rho})$ from eq.~(\ref{f6}) into (\ref{f5}) results in the closed nonlinear equation for electron wave function $\chi(\bm{\rho}')$. The first term in (\ref{f6}) leads to an additional potential affecting the electron and caused by that part of the fluctuation density in the exciton condensate which appears due to the interaction with a remote ionized donor. This contribution can be essentially simplified and expressed by a single integral:
\begin{eqnarray}\label{f7}
&&\frac{1}{2m}\Delta_{\bm{\rho}'}\chi(\bm{\rho}')+\left(E+2\pi\tilde{e}^2Ln_0\right)\chi(\bm{\rho}')-U_0(\bm{\rho}')\chi(\bm{\rho}')-\nonumber\\
&&-\frac{4\pi\tilde{e}^2n_0L^2}{a_M}\int\frac{e^{-k(\Delta+\Delta_1)}J_0(k\rho')kdk}{\kappa^2+k^2}\chi(\bm{\rho}')+\nonumber\\
&&+\frac{2Mn_0(\tilde{e}^2\Delta L)^2}{\pi}\chi(\bm{\rho}')\times\\\nonumber
\\\nonumber
&&\times\int\frac{K_0(\kappa|\bm{\rho}_1-\bm{\rho}_2|)\left|\chi(\bm{\rho}_3)\right|^2d\bm{\rho}_1d\bm{\rho}_2d\bm{\rho}_3}{\left[(\bm{\rho}_2-\bm{\rho}_3)^2+\Delta^2\right]^{3/2}\left[(\bm{\rho}'-\bm{\rho}_1)^2+\Delta^2\right]^{3/2}}=0,\nonumber\\\nonumber
\end{eqnarray}
where $a_M=1/M\tilde{e}^2$ (Bohr radius for a particle with mass $M$).

To approximately find the eigenvalues of eq.~(\ref{f7}), which determine the electron energy levels in a QD, we use the direct variation method. We model the QD plus ionized donor potential $U_0(\bm{\rho})$ by the parabolic one: $U_0(\bm{\rho})=m\Omega^2\rho^2/2$. Correspondingly, we choose the trial functions of the lowest and of the first excited levels of QD as the eigenfunctions of a 2D harmonic oscillator:
\begin{eqnarray}\label{f8}
\chi_0=\sqrt{\frac{\alpha}{\pi}}e^{-\alpha\rho'^2/2}, ~~~\chi_1=\sqrt{\frac{2}{\pi}}\beta\rho'\cos\varphi e^{-\beta\rho'^2/2},
\end{eqnarray}
where $\alpha$ and $\beta$ are the variation parameters. Functions (\ref{f8}) are orthogonal for any $\alpha$ and $\beta$ and normalized.

Before calculating energy levels $E_0$ and $E_1$, we discuss the condition of applicability of the linear in $\varphi$ approximation used ($\varphi\ll\sqrt{n_0}$). To this end, calculate $\varphi_(\bm{\rho})$ in (\ref{f6}) with the function $\chi_0$ at point $\rho=0$ (i.e. just under the QD). Put $\chi_0$ from eq.~(\ref{f8}) into (\ref{f6}) and replace function $K_0$ by its Fourier transform. The result has the form:
\begin{eqnarray}\label{f9}
\varphi(0)=\frac{2L\sqrt{n_0}}{a_M}\int\left(e^{-k^2/4\alpha-k\Delta}-e^{-k\Delta_1}\right)\frac{kdk}{\kappa^2+k^2}.
\end{eqnarray}

As $\Delta_1>\Delta$ (see Fig.~\ref{fig1}), then, at $\alpha\gg 1/\Delta^2$ (that means the lateral size of QD is much smaller than $\Delta$), the integrand in (\ref{f9}) is positive, and it is less than $ke^{-k\Delta}/\kappa^2$ everywhere. Hence, $\varphi(0)<2L\sqrt{n_0}/a_M\kappa^2\Delta^2$. The required criterion takes the form of $8\pi n_0\Delta^2\gg1$. In the opposite limit $\alpha\Delta^2\ll1$, the second term in the round bracket of integrand in (\ref{f9}) dominates, value $\varphi(0)$ changes the sign and its modulus is less than $2L\sqrt{n_0}/a_M\kappa^2\Delta_1^2$. Then $|\varphi(0)|\ll\sqrt{n_0}$ if $8\pi n_0\Delta_1^2\gg1$. Thus, the previous condition $8\pi n_0\Delta^2\gg1$ provides the linearization validity of system (\ref{f3}) for all QD sizes.

The energies of the two lowest levels as the functions of variational parameters $E_0(\alpha)$ and $E_1(\beta)$ are found by the conventional method of substitution $\chi_0$ and $\chi_1$ in the functional of the energy related to eq.~(\ref{f7}). This functional, as usually, is average value of the total energy $\langle\Psi|\hat{T}+U_{eff}|\Psi\rangle$ where $\hat{T}$ is the kinetic energy operator and $U_{eff}$ is the effective potential energy of the electron in QD which is given by all terms in the eq.~(\ref{f7}) except the first one. All integrations over coordinates $\bm{\rho}_1$, $\bm{\rho}_2$, etc. can be done analytically and the final results contain single integrals only (we count energy from the general shift $-2\pi\tilde{e}^2Ln_0$ which results from the interaction with the background uniform density of the excitons):
\begin{eqnarray}
&&E_0(\alpha)=\frac12\left(\frac{\alpha}{m}+\frac{m\Omega^2}{\alpha}\right)-\nonumber\\
&&-\frac{8\pi n_0L^2\tilde{e}^2}{a_M}\int\limits_0^{\infty}\frac{e^{-k^2/2\alpha-2k\Delta}kdk}{\kappa^2+k^2}+\nonumber\\
&&+\frac{4\pi n_0L^2\tilde{e}^2}{a_M}\int\limits_0^{\infty}\frac{e^{-k^2/4\alpha-k(\Delta+\Delta_1)}kdk}{\kappa^2+k^2},\label{f10}\\
&&E_1(\beta)=\frac{\beta}{m}+\frac{m\Omega^2}{\beta}-\nonumber\\
&&-\frac{8\pi n_0L^2\tilde{e}^2}{a_M}\int\limits_0^{\infty}\frac{e^{-k^2/2\beta-2k\Delta}}{\kappa^2+k^2}\left(1-\frac{k^2}{2\beta}+\frac{3k^4}{32\beta^2}\right)kdk+\nonumber\\
&&+\frac{4\pi n_0L^2\tilde{e}^2}{a_M}\int\limits_0^{\infty}\left(1-\frac{k^2}{4\beta}\right)\frac{e^{-k^2/4\beta-k(\Delta+\Delta_1)}kdk}{\kappa^2+k^2}.\label{f11}
\end{eqnarray}

\begin{figure}
\onefigure[width=5.5cm]{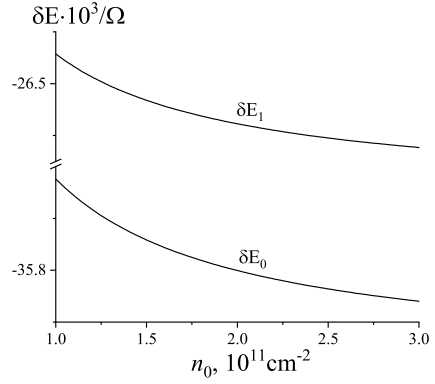}
\caption{Density dependence of the first excited (upper panel) and the ground (lower panel) QD levels shifts ($\delta E_0=E_0-\Omega$ and $\delta E_1=E_1-2\Omega$) for the condensate phase of excitons at $T=0$.}
\label{fig2}
\end{figure}

Equations $\partial E_0/\partial\alpha=0$, $\partial E_1/\partial\beta=0$ have been numerically solved; the values $E_0$ and $E_1$ as the functions of condensate density $n_0$ are plotted in Fig.~\ref{fig2}. The unperturbed values of these levels, evidently, are: $E_0=\Omega$, $E_1=2\Omega$. Calculations were performed with parameters $\varepsilon=12.5$, $\Delta=20$~nm, $\Delta_1=25$~nm, $L=6$~nm, $\Omega=1.9$~meV, $M=0.6m_0$, $m=0.1m_0$, where $m_0$ is the electron mass.

\section{Normal Bose-gas of excitons}

In this case we describe excitons quasiclassically in the self-consistent field approximation. The Hamiltonian of the electron-exciton interaction can be expressed through the local exciton density $n(\bm{\rho})$:
\begin{eqnarray}\label{f12}
\hat{H}_{int}=\int|\chi(\bm{\rho}')|^2V_{\mbox{e-ex}}(\bm{\rho}'-\bm{\rho})n(\bm{\rho})d\bm{\rho}'d\bm{\rho}.
\end{eqnarray}

The density $n(\bm{\rho})$ above the condensation point is determined by the Bose-Einstein distribution accounting for the potential $W(\bm{\rho})$ - the energy of a dipole in the field of all other excitons and interaction with the electron in a QD and with an ionized donor:
\begin{eqnarray}\label{f13}
W(\bm{\rho})=gn(\bm{\rho})+\int V_{\mbox{e-ex}}(\bm{\rho}-\bm{\rho}')|\chi(\bm{\rho}')|^2d\bm{\rho}'+\nonumber\\
+\frac{\tilde{e}^2L\Delta_1}{\left(\Delta_1^2+\rho^2\right)^{3/2}}.
\end{eqnarray}

The exciton density in the quasiclassical approach is determined by the formula:
\begin{eqnarray}\label{f14}
n(\bm{\rho})=\frac{1}{(2\pi)^2}\int\frac{d\bm{p}}{e^{\beta[p^2/2M+W(\bm{\rho})-\mu]}-1}=\nonumber\\
-\frac{M}{2\pi\beta}\ln\left[1-e^{\beta[\mu-W(\bm{\rho})]}\right],~~~\beta\equiv1/T.
\end{eqnarray}

In the absence of QD and donor (but with exciton-exciton interaction allowed for), we have $n=n_0$ - the density of the unperturbed exciton gas, $W=gn_0$. Consider now the Boltzmann limit that occurs at $2\pi\beta n(\bm{\rho})/M\ll 1$. Then logarithm in eq.~(\ref{f14}) can be expanded and we come to the barometric formula for the exciton density: $n=n_0e^{-\beta\tilde{W}}$, where $\tilde{W}=W-gn_0$. Estimating $M=0.6m_0$, $n_0\sim10^{11}\mbox{cm}^{-2}$, we see that this limit is reached at $T\gg T_B=2\pi n_0/M\sim7$~K. The simplest case is the one of high temperature $\beta\tilde{W}\ll1$. As follows from eq.~(\ref{f13}), this condition is met when $T\gg|V_{\mbox{e-ex}}|$, $V_{\mbox{D-ex}}$, i.e. it is enough to use $\tilde{e}^2L/\Delta^2\ll T$. For example, if the dipole shoulder $L$  in a double quantum well equals 6~nm, $\Delta\sim30$~nm and $\varepsilon=12.5$, $\tilde{e}^2L/\Delta^2\approx7.6$~K is obtained and the Boltzmann approximation $T\gg T_B$  provides the condition $\tilde{W}\ll T$, too. Then the deviation of the local density from its equilibrium value $n_0$ is proportional to the potential  $\tilde{W}$ that creates this deviation (quite similar to the linear regime in screening). Then $n=n_0(1-\beta\tilde{W})$ and, from (\ref{f13}), we have
\begin{eqnarray}\label{f15}
\tilde{W}(\bm{\rho})(1+\beta gn_0)=\int V_{\mbox{e-ex}}(\bm{\rho}-\bm{\rho}')|\chi(\bm{\rho}')|^2d\bm{\rho}'+V_{\mbox{D-ex}}(\bm{\rho}).\nonumber\\
\end{eqnarray}

As $n-n_0=-n_0\beta\tilde{W}$ we find from Eqs.~(\ref{f12}) and (\ref{f15}) additional potential energy of the electron in QD caused by the density perturbation in the exciton gas and finally, we come to the Schr\"{o}dinger-type equation for the electron wave function:
\begin{eqnarray}\label{f16}
&&-\frac{1}{2m}\Delta\chi(\bm{\rho}')+\Bigg\{\frac{m\Omega^2\rho'^2}{2}-\frac{n_0}{T+gn_0}\int V_{\mbox{e-ex}}(\bm{\rho}'-\bm{\rho}_1)\times\nonumber\\
&&\times\Big[V_{\mbox{e-ex}}(\bm{\rho}_1-\bm{\rho}_2)|\chi(\bm{\rho}_2)|^2d\bm{\rho}_2+\\
&&~~~~~~~~~~~~~~~~~~+V_{\mbox{D-ex}}(\bm{\rho}_1)\Big]d\bm{\rho}_1\Bigg\}\chi(\bm{\rho}')=E\chi(\bm{\rho}').\nonumber
\end{eqnarray}

With the same trial functions $\chi_0$ and $\chi_1$, for the levels $E_0$ and $E_1$, in the case of the normal state of the exciton gas, we obtain:
\begin{eqnarray}
&&E_0(\alpha)=\frac12\left(\frac{\alpha}{m}+\frac{m\Omega^2}{\alpha}\right)-\nonumber\\
&&-\frac{2\pi n_0L^2\tilde{e}^4}{T+gn_0}\int\limits_0^{\infty}e^{-k^2/2\alpha-2k\Delta}kdk+\label{f17}\\
&&+\frac{4\pi(\Delta+\Delta_1)\alpha n_0L^2\tilde{e}^4}{T+gn_0}\int\limits_0^{\infty}\frac{e^{-\alpha\rho^2}\rho d\rho}{\left[(\Delta+\Delta_1)^2+\rho^2\right]^{3/2}},\nonumber\\
&&E_1(\beta)=\frac{\beta}{m}+\frac{m\Omega^2}{\beta}-\nonumber\\
&&-\frac{2\pi n_0L^2\tilde{e}^4}{T+gn_0}\int\limits_0^{\infty}e^{-k^2/2\beta-2k\Delta}\left(1-\frac{k^2}{2\beta}+\frac{3k^4}{32\beta^2}\right)kdk+\nonumber\\
&&+\frac{4\pi(\Delta+\Delta_1)\beta^2n_0L^2\tilde{e}^4}{T+gn_0}\int\limits_0^{\infty}\frac{e^{-\beta\rho^2}\rho^3d\rho}{\left[(\Delta+\Delta_1)^2+\rho^2\right]^{3/2}}.\label{f18}
\end{eqnarray}

\begin{figure}
\onefigure[width=5.5cm]{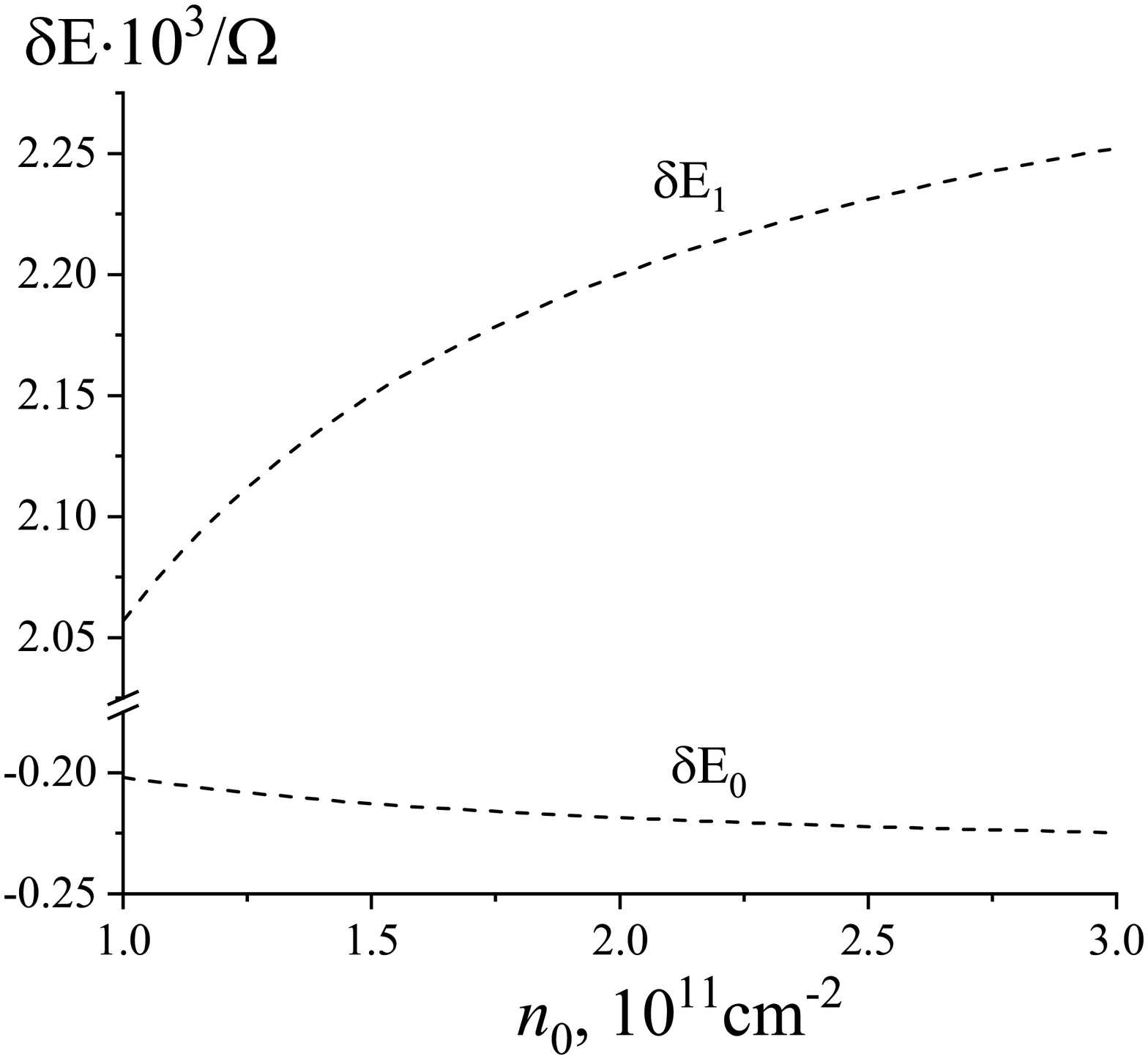}
\caption{Same levels for non-degenerate exciton gas at $T=15$~K.}
\label{fig3}
\end{figure}

All integrals in (\ref{f17}) and (\ref{f18}) can be reduced to the function $\Phi(x)$ - the probability integral, but the formulae become too cumbersome. The numerically calculated $E_0$ and $E_1$ are shown in Fig.~(\ref{fig3}) for $T=15$~K.
\begin{figure}
\onefigure[width=5.5cm]{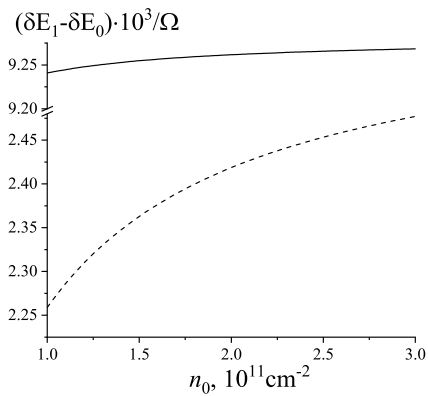}
\caption{$(\delta E_1-\delta E_0)/\Omega$ for $T=0$ (solid line) and $T=15$~K (dashed line).}
\label{fig4}
\end{figure}
\begin{figure}
\onefigure[width=5.5cm]{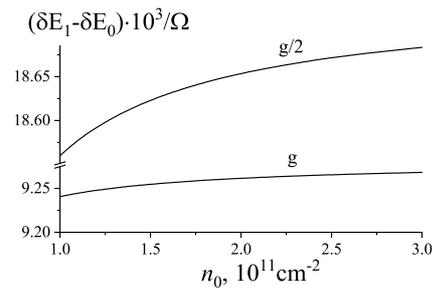}
\caption{$(\delta E_1-\delta E_0)/\Omega$ at $T=0$ for $g$ and $g/2$.}
\label{fig5}
\end{figure}
It is instructive to trace how parameters $\Delta$ and $\Delta_1$ enter the last terms of equations (\ref{f10}),(\ref{f11}) and (\ref{f17}),(\ref{f18}) describing the remote donor contribution to the energy shift of the electron levels in a QD. Direct Coulomb electron-donor interaction is included in $U_0$. Thus, it goes about an effect mediated by the changes in the exciton density. The donor-exciton interaction (distance $\Delta_1$) causes the exciton density variation, which, in its turn, affects the QD electron (distance $\Delta$). As a result, the final expression contains the sum $\Delta + \Delta_1$.

Thus we see that the shifts of the electron energy levels in QD are essentially different in the condensate phase and in the normal (Boltzmann) bose gas. In the latter case even sign of the shift becomes positive for the level $E_1$ because of weaker screening of the donor potential in the state $E_1$ than in the state $E_0$: the wave function $\chi_0$ gives maximum of the electron charge exactly ''under'' the donor charge. In general the BEC state results in larger levels shift that the normal state.

To conclude, we have shown that the phase transition normal Bose-gas $\to$ BEC in the hybrid structure containing QDs and the 2D gas of dipolar excitons can be detected by the measurements of electron transitions in a QD.

Such experiments seem difficult because the frequency shifts are rather small. However raman and photoluminescence spectroscopy of single QDs demonstrate  very high accuracy. E.g. in the works \cite{a18,a19} with GaAs single QDs authors show five  experimental points between 1.6229 meV and 1.6230 meV. Such accuracy would be enough, as follows from the Fig.~4, for experiments with Ge QDs (see ref.~\cite{a17}).

Next issue relates to the approximation of a weakly non-ideal bose gas that ignores correlation effects. These effects were discussed in the literature \cite{a20,a21,a22}, and it has been shown that renormalization of the bare interaction between the dipoles can result in strong decrease in the effective coupling constant $g$. Just to illustrate the influence of change in $g$ on QD electron spectrum we have repeated our computations for $g$ two times smaller than for the plane capasitor model used above. Results are shown in the Fig.~5. As expected the shift of the electron transition frequency in QD strongly depends on the coupling constant $g$. Decrease in the magnitude of $g$ makes the above discussed effect more pronounced ($g$ stands in the denominator of the integrand in eq.~(\ref{f7}): $\kappa^2\sim g$). Hence, the computations made in this work with plane capacitor model for $g$ give the minimal estimate of the effect in question. Dependence of the coupling constant on the density $n_0$ is a separate and rather interesting problem but this is beyond the framework of the present paper.

\acknowledgments
This research was supported in part by the RFBR, grant No. 20-02-00622.

\end{document}